\documentclass{PoS}

\usepackage{epsfig}

\title{2+1 flavor light hadron spectrum and quark masses
       with the $O(a)$ improved Wilson-clover quark formalism}

\ShortTitle{2+1 flavor light hadron spectrum and quark masses}

\author{CP-PACS and JLQCD Collaborations:}

\author{\speaker{T.~Ishikawa}
        \thanks{E-mail: tomomi@ccs.tsukuba.ac.jp}~${}~^a$,
	S.~Aoki${}^{a,~b}$, M.~Fukugita${}^c$,
	S.~Hashimoto${}^{d,~e}$, K.-I.~Ishikawa${}^f$, N.~Ishizuka${}^{a,~g}$,
	Y.~Iwasaki${}^a$, K.~Kanaya${}^a$, T.~Kaneko${}^{d,~e}$,
	Y.~Kuramashi${}^{a,~g}$, M.~Okawa${}^f$, Y.~Taniguchi${}^{a,~g}$,
	N.~Tsutsui${}^d$, A.~Ukawa${}^{a,~g}$, N.~Yamada${}^{d,~e}$
	and T.~Yoshi\'{e}${}^{a,~g}$\\

	${}^a$ Graduate School of Pure and Applied Sciences,
	       University of Tsukuba,
	       Tsukuba, Ibaraki 305-8571, Japan\\
	${}^b$ Riken BNL Research Center, Brookhaven National Laboratory,
	       Upton, New York 11973, USA\\
	${}^c$ Institute for Cosmic Ray Research, University of Tokyo,
               Kashiwa 277-8582, Japan\\
        ${}^d$ High Energy Accelerator Research Organization (KEK),
               Tsukuba 305-0801, Japan\\
	${}^e$ School of High Energy Accelerator Science,
  	       The Graduate University for Advanced Studies (Sokendai),
  	       Tsukuba 305-0801, Japan\\
        ${}^f$ Department of Physics, Hiroshima University,
               Higashi-Hiroshima, Hiroshima 739-8526, Japan\\
	${}^g$ Center for Computational Sciences, University of Tsukuba,
 	       Tsukuba, Ibaraki 305-8577, Japan
	}

\abstract{
We present a summary of results of the joint CP-PACS and JLQCD project 
toward a 2+1 flavor full QCD simulation with the $O(a)$-improved
Wilson quark formalism and the Iwasaki gauge action. 
Configurations were generated during 2002--2005 at three lattice spacings,
$a\sim 0.076$, $0.100$ and $0.122$~fm,
keeping the physical volume constant at $(2.0~\mbox{fm})^3$.
Up and down quark masses are taken in the range $m_{PS}/m_V\sim0.6-0.78$.
We have completed the analysis for the light meson spectrum 
and quark masses in the continuum limit using the 
full configuration set.  The predicted meson masses reproduce experimental 
values in the continuum limit at a 1\% level. 
The average up and down, and strange quark masses turn out to be
$m_{ud}^{\overline{\rm MS}}(\mu=2~{\rm GeV})
=3.50(14)({}^{+26}_{-15})~{\rm MeV}$
and
$m_s^{\overline{\rm MS}}(\mu=2~{\rm GeV})
=91.8(3.9)({}^{+6.8}_{-4.1})~{\rm MeV}$.
We discuss our future strategy toward definitive results on hadron 
spectroscopy with the Wilson-clover formalism. }

\FullConference{XXIV International Symposium on Lattice Field Theory\\
                 July 23-28 2006\\
                 Tucson Arizona, US}

\begin{document}

\begin{table}[b]
\begin{center}
\begin{tabular}{ccccc}
 \hline
 $\beta$ & size & $a$ [fm] ($K$-input) & $a$ [fm] ($\phi$-input) &
 trajectory \\
 \hline
 $1.83$ & $16^3\times32$ & $0.1222(17)$ & $0.1233(20)$ & $7000 - 8600$ \\
 $1.90$ & $20^3\times40$ & $0.0994(19)$ & $0.0995(19)$ & $5000 - 9200$ \\
 $2.05$ & $28^3\times56$ & $0.0693(26)$ & $0.0695(26)$ & $6000 - 6500$ \\
 \hline
\end{tabular}
\caption{Main simulation parameters.}
\label{TAB:simulation_parameters}
\end{center}
\end{table}

\section{Introduction}

The calculation of the light hadron spectrum and quark masses is 
a fundamental and necessary step for the entire area of lattice QCD 
simulations. 
Moving from a precision quenched calculation~\cite{Aoki:2000-1} to two flavor 
($N_f=2$) full QCD~\cite{Ali_Khan:2000-1}, we observed 
a significant dynamical up and down  (``light'') quark effect,
which removes most of the O(10\%) systematic deviation in  
the quenched QCD spectrum from experiment.  It was also found that 
light quark masses are significantly reduced in $N_f=2$ full QCD. 
In order to remove the quenching error of the heavier strange quark,
the CP-PACS and JLQCD collaborations have jointly pursued a 2+1 flavor
($N_f=2+1$) full QCD 
simulation~\cite{KanekoTIshikawa:2004-1:2005-1:2005-2} since 2001.
We employ the Wilson quark formalism as in our quenched and $N_f=2$ studies,
preferring an unambiguous quark-flavor interpretation over the computational 
ease of the staggered formalism carried out by the MILC 
collaboration~\cite{Aubin:2004-1}.

The project explored the light quark mass range corresponding to 
$m_\pi/m_\rho=0.6-0.78$ for which the configuration generation has been 
completed at three lattice spacings in the fall of 2005.
In this article, we present a summary of the results for light meson masses, 
light quark masses, pseudoscalar (PS) decay constants and the Sommer scale,
evaluated in the continuum limit using the full set of configurations. 
We also discuss systematic error from chiral extrapolations 
by comparing fits with polynomial functions of quark masses
and those based on chiral perturbation theory ($\chi$PT).

\section{Production of the gauge configuration}
For the lattice action, we employ the renormalization group
(RG) improved Iwasaki gauge action and the clover quark action
with the improvement coefficient $c_{SW}$ determined 
non-perturbatively for the RG action~\cite{Aoki:2005-1}.
The choice of the gauge action is made to avoid a first-order phase
transition (lattice artifact) observed for the plaquette gauge 
action \cite{Aoki:2004-bulk}.

Configurations are generated with the Polynomial Hybrid Monte Carlo
(PHMC) algorithm.
(See \cite{Aoki:2002-2} for our implementation.) 
The molecular dynamics time step $\delta\tau$ 
and the polynomial order $N_{poly}$ are chosen such that 
the HMC and the global Metropolis acceptance rate achieves 
$85\%$ and $90\%$, respectively.

Simulations are performed at three values of the coupling constant
such that $a^2$ is placed at an even interval.
The physical volume is fixed at $(2.0~\mbox{fm})^3$. 
Main simulation parameters are listed in 
Table~\ref{TAB:simulation_parameters}.
At each coupling, we generate configurations for ten combinations
of hopping parameters $(\kappa_{ud},\kappa_s)$, five for 
the ud quark mass taken in the range of the pseudoscalar (PS) to
vector (V) meson mass ratio of $m_{PS}/m_V\sim 0.6-0.78$ 
and two for the strange quark mass chosen around $m_{PS}/m_V\sim 0.7$.

\section{Measurement and analysis}\label{SEC:measurement}

Measurements are made at every 10 HMC trajectories.
We use the combination of smeared source and point sink,
with which the effective masses reach  plateau
earliest and the statistical errors are smallest.
Meson masses and amplitude of the correlator are obtained from
single mass $\chi^2$ fits to correlators $\langle P(t)P(0)\rangle$, 
$\langle V(t)V(0)\rangle$ and $\langle A_4(t)P(0)\rangle$, 
where $P$, $V$ and $A_{\mu}$ denote the PS, 
the vector and the non-perturbatively $O(a)$-improved~\cite{Kaneko:2006-1}
axial-vector current, respectively. 
We include correlations in time but ignore those among correlators,
since our statistics are not sufficient for the latter. 
Errors are estimated by the binned jackknife method
with the bin size of $100$ HMC trajectories. 

For chiral extrapolations of meson masses, we use two definitions 
of the quark mass, 
the vector Ward identity (VWI) quark mass 
$m_q^{VWI}=(1/\kappa-1/\kappa_c)/2$ ($\kappa_c$ is the critical 
hopping parameter where $m_{PS}$ at 
$\kappa_{ud}=\kappa_s=\kappa_{val}=\kappa_c$ vanishes), and
the axial-vector Ward identity (AWI) quark mass
$m_q^{AWI}=\nabla_{\mu}A_{\mu}(x)/(2P(x))$.
Quark masses are determined from chiral fits with $m_q^{AWI}$,
because $m_q^{VWI}$ shows large scaling violation.
\footnote{
The VWI quark mass for the ud quarks is negative at our simulation 
points. This originates from a lack of chiral symmetry of the Wilson
quark action. This is another reason to prefer the AWI definition.}
We use chiral fits with $m_q^{VWI}$ for other quantities. 

Chiral fits are made to light-light, light-strange and
strange-strange meson masses simultaneously ignoring their correlations,
using a quadratic polynomial function of the sea and valence quark
masses;
\begin{eqnarray}
f({\rm M_s}, {\rm M_v})&=&A+B_S{\rm tr}{\rm M_s}+B_V{\rm tr}{\rm M_v}
+D_{SV}{\rm tr}{\rm M_s}{\rm tr}{\rm M_v}\nonumber\\
&&+C_{S1}{\rm tr}{\rm M_s^2}+C_{S2}({\rm tr}{\rm M_s})^2
+C_{V1}{\rm tr}{\rm M_v^2}+C_{V2}({\rm tr}{\rm M_v})^2,
\end{eqnarray}
where ${\rm M_S}={\rm diag}(m_{ud}, m_{ud}, m_s)$,
${\rm M_V}={\rm diag}(m_{val 1}, m_{val 2})$,
and ``${\rm tr}$'' means trace of matrices.
We set $A=0$ for fits of $m_{PS}^2$ with $m_q^{VWI}$, while
$A=B_S=C_{S1}=C_{S2}=0$ for those with $m_q^{AWI}$.

The physical point is fixed for two cases.
The ``$K$-input'' takes the experimental values  
$m_{\pi}=0.1350~\mbox{GeV}$, $m_{\rho}=0.7684~\mbox{GeV}$
and $m_K=0.4977~\mbox{GeV}$ as inputs.
In the ``$\phi$-input'' case, $m_{\pi}$, $m_{\rho}$ and
$m_{\phi}=1.0194~\mbox{GeV}$ are taken as inputs.
The lattice spacings determined from
the $K$- and $\phi$- inputs are consistent with each other,
as shown in Table~\ref{TAB:simulation_parameters}.

Static potential $V(r)$ is determined from Wilson loops for 
smeared gauge links and is fitted to a form 
$V(r)=c-\alpha/r + \sigma r$.
The Sommer scale $r_0$ calculated from the fit is extrapolated
to the physical point as $1/r_0=A+B_S{\rm tr}{\rm M_s}$ using
$m_q^{VWI}$.

\section{Physics results}

\subsection{Light meson spectrum}

\begin{figure}
\begin{center}
\includegraphics[scale=0.55, viewport = 0 0 750 360, clip]
 {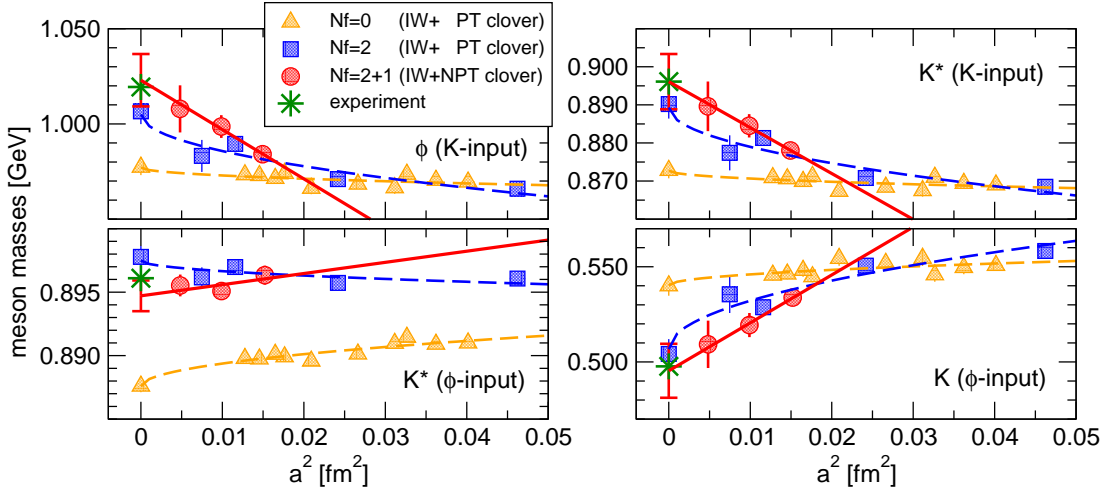}
\vspace*{-5mm}
\caption{Continuum extrapolation of meson masses,
         comparing with those for quenched and $N_f=2$ 
         QCD~\protect\cite{Ali_Khan:2000-1}.
         Note that the quenched and $N_f=2$ simulations are made with 
         the one-loop perturbatively $O(a)$-improved clover action.
         Thus extrapolations are made linearly in $a$.}
\label{FIG:meson_masses}
\end{center}
\vspace*{-5mm}
\end{figure}

The meson masses are well fitted by linear functions in $a^2$, as shown in
Fig.~\ref{FIG:meson_masses}. We obtain in the continuum limit, 
\begin{eqnarray}
&&m_{K^{\ast}}=0.8961(72)~{\rm GeV},\;\;\;\;
m_{\phi}=1.023(14)~{\rm GeV}\;\;\;\;\;\;\;\;(K-{\rm input}),\nonumber\\
&&\;m_K=0.495(14)~{\rm GeV},\;\;\;\;\;\;
m_{K^{\ast}}=0.8947(12)~{\rm GeV}\;\;\;\;\;(\phi-{\rm input}).
\end{eqnarray}
These values are consistent with experiment within the quoted 
statistical errors of about 1\% level.  The errors are larger than those 
achieved in the $N_f=2$ case, and possible deviations from $N_f=2$ QCD 
are not detected. 
We observe that the scaling violation in $N_f=2+1$ QCD seems to be larger 
than that in the quenched or $N_f=2$ case~\cite{Ali_Khan:2000-1}
even though the actions used in these cases are less improved than the 
present $N_f=2+1$ case. 
We may argue, however, that the magnitude of the scaling violation in
$N_f=2+1$ QCD is not particularly large. A quadratic fit of 
the $K^*$ meson mass with $K$-input $m_{K^*}=m_0(1+c(\Lambda_{QCD}\cdot a)^2)$
yields $m_0=8960(73)$~MeV and $c=-1.29(59)$ for $\Lambda_{QCD}=200$~MeV.
The O(1) magnitude of the coefficient $c$ is reasonable.

\subsection{Quark masses}

The physical quark mass is determined for the $\overline{\mbox{MS}}$ 
scheme at the scale $\mu=2$~GeV. 
Lattice results are translated to the $\overline{\mbox{MS}}$ scheme
at $\mu=a^{-1}$ using tadpole-improved one-loop matching 
factor~\cite{Aoki:1998-PT-renorm}, and then evolved to 
$\mu=2$~GeV using the four-loop RG-equation.

\begin{figure}
\begin{center}
\includegraphics[scale=0.52, viewport = 0 0 750 270, clip]
 {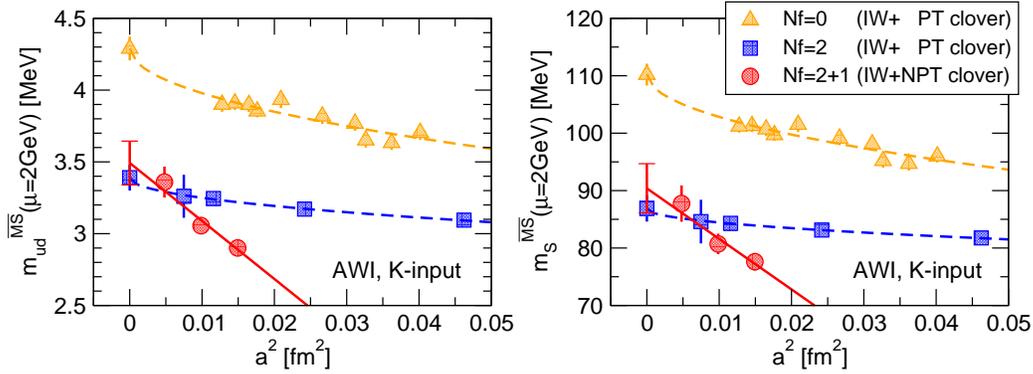}
\caption{Continuum extrapolations of the up, down and strange quark masses
         obtained with the $K$-input.
         For comparison, results for quenched and $N_f=2$ QCD
         ~\protect\cite{Ali_Khan:2000-1} are overlaid.}
\label{FIG:uds_quark_masses}
\end{center}
\vspace*{-5mm}
\end{figure}

Quark mass results are shown in Fig.~\ref{FIG:uds_quark_masses}.
While $O(g^4a)$ scaling violation should be present due to the use of 
one-loop matching factor, comparison of VWI and AWI masses for ud quarks 
suggests that these terms are small relative to the $O(a^2)$
term~\cite{KanekoTIshikawa:2004-1:2005-1:2005-2}.
Therefore we extrapolate quark masses linearly in $a^2$. 
Possible effects of $O(g^4a)$ terms are estimated from the 
ambiguity of the renormalization factor 
by either shifting the matching scale from $\mu=1/a$ to 
$\mu=\pi/a$ or using an alternative definition of coupling for
tadpole improvement~\cite{Ali_Khan:2000-1}. 
  
As already observed in $N_f=2$ QCD~\cite{Ali_Khan:2000-1}, 
values of the strange quark mass determined for either the $K$- or 
the $\phi$-inputs, while different at finite lattice spacings, extrapolate 
to a common value in the continuum limit. 
Therefore the quark masses in the continuum limit is estimated from 
a combined fit to data with the $K$- and the $\phi$-inputs.
We finally obtain
\begin{equation}
 m_{ud}^{\overline{\rm MS}}(\mu=2~{\rm GeV})
=3.50(14)({}^{+26}_{-15})~{\rm MeV},\;\;\;\;\;
 m_s^{\overline{\rm MS}}(\mu
=2~{\rm GeV})=91.8(3.9)({}^{+6.8}_{-4.1})~{\rm MeV}.
\end{equation}
Dynamical up and down quarks reduce significantly the quark 
masses~\cite{Ali_Khan:2000-1}.
The effect of strange quark is less dramatic, and we do  not see 
deviations from the $N_f=2$ results beyond statistical errors.

\subsection{PS decay constants}

PS meson decay constants are estimated using matching factor
determined by tadpole-improved one-loop perturbation theory.
The results with $K$-input are
\begin{equation}
f_{\pi}=143.4(8.8)~{\rm GeV},\;\;\;
f_K=163.7(8.6)~{\rm GeV},\;\;\;
f_K/f_{\pi}=1.140(17).
\end{equation}
We recall that in our $N_f=2$ QCD calculation, 
the magnitude of scaling violation was so large that we were not able 
to estimate values in the continuum 
limit~\cite{Ali_Khan:2000-1}.
The situation is much better in the present case and $f_{\pi}$ and $f_K$ turn
out to be almost consistent with experiment.
The errors are large, however.  
Furthermore, the ratio $f_K/f_{\pi}$ differs significantly from experiment.
A long chiral extrapolation is a possible cause of the discrepancy. 

\subsection{Sommer scale}

The Sommer scale in the continuum limit for $K$-input reads
\begin{equation}
r_0=0.516(21)~{\rm fm},
\end{equation}
which is closer to a phenomenological value of $0.5$ fm than
the estimate $r_0\approx 0.541(17)$ fm in the $N_f=2$ 
QCD~\cite{Ali_Khan:2000-1}. 

\section{Chiral fit using the Wilson $\chi$PT}\label{SEC:WChPT}

Our long chiral extrapolation is a dangerous source of systematic error.
As a supplementary analysis, we try to fit meson masses using the $\chi$PT
and compare it with the polynomial fit above.

We use the $\chi$PT modified for the Wilson quark action
(W$\chi$PT)~\cite{Sharpe:1998-1}, in which 
finite lattice spacing corrections to infrared chiral logarithms
for the Wilson quark action are incorporated.
The $N_f=2+1$ QCD W$\chi$PT formulae for the $O(a)$ improved theory
has been obtained~\cite{Aoki:2003-WChPT_Nf2} up to the NLO. 
For our fit, we rewrite the formulae in terms of the AWI quark mass; 
\begin{eqnarray}
m_{\theta}^2&=&(x+P_{\theta}y)\Biggl[1+\frac{1}{f^2}\Biggl\{
\sum_{\psi=\pi,K,\eta}L_{\psi}(\mu)A_{\psi}^{\theta}
-H-12L_{46}(\mu)x-4L_{58}(\mu)(x+P_{\theta}y)\Biggr\}\Biggr],
\label{EQ:WChPT-PS}\\
m_{\omega}&=&m_O+\lambda_xx+\lambda_yy-\frac{1}{24\pi^2f^2}
\sum_{\psi=\pi,K,\eta}H_{\psi}^{\omega}(g_1^2, g_2^2)(x+P_{\psi}y)^{3/2},
\label{EQ:WChPT-V}
\end{eqnarray}
where $\theta=\pi, K$ and $\omega=\rho, K^{\ast}$.
In these equations $P_{\theta}, A_{\psi}^{\theta}$ and
$H_{\psi}^{\omega}(g_1^2, g_2^2)$ are geometric factors,
$f$ is the pion decay constant in the chiral limit of the LO formula, and
\begin{equation}
x=\frac{2B_0}{3}(2m_{ud}+m_s),\;\;\; y=\frac{B_0}{3}(m_{ud}-m_s),
\end{equation}
\begin{equation}
L_{46}(\mu)=2L_4(\mu)+4L_6(\mu),\;\;\; L_{58}(\mu)=2L_5(\mu)+4L_8(\mu),
\end{equation}
\begin{equation}
L_{\psi}(\mu)=\frac{x+P_{\psi}y}{16\pi^2}\ln\frac{x+P_{\psi}y}{\mu^2},
\end{equation}
where $B_0$, $L_4$, $L_5$, $L_6$ and $L_8$ are the low energy constants
in the continuum $\chi$PT.
We note that up to NLO the vector formula has no lattice artifact
whereas the PS formula has the term $H$ which is of $O(a^2)$.

\begin{figure}[t]
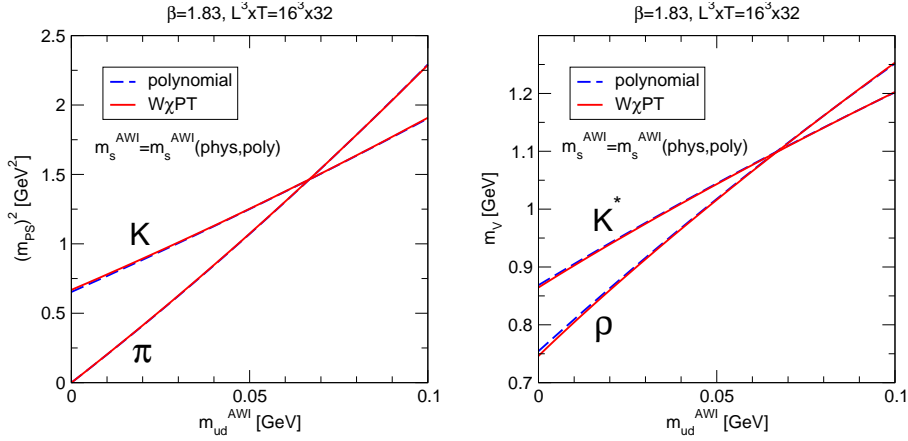

\begin{center}
\parbox{60mm}{
\includegraphics[scale=0.40, viewport = 0 0 410 410, clip]
                 {Figures/wchpt_mps2_B1.83.eps}
}
\hspace*{+1mm}
\parbox{60mm}{
\includegraphics[scale=0.40, viewport = 0 0 410 410, clip]
                 {Figures/wchpt_mv_B1.83.eps}
}
\caption{
Comparison of the W$\chi$PT fit and the polynomial chiral fits
at $\beta=1.83$.
$m_s^{AWI}$ and the lattice spacing necessary to draw curves are
set by polynomial chiral fits and the $K$-input.}
\label{FIG:Chiral_fit_WChPT}
\end{center}
\vspace*{-5mm}
\end{figure}

We find that while fit parameters vary largely among jackknife samples, 
and hence are not well determined,
fitting curves themselves are stable.
Also the W$\chi$PT fitting curves do not exhibit any significant difference
from the polynomial fit, as shown in Fig.~\ref{FIG:Chiral_fit_WChPT}. 
Therefore, conclusions for the spectrum and quark masses derived from
polynomial fits are not altered even if we use the W$\chi$PT fit.
We suppose that a relatively large value of the up and down quark
masses in our simulation makes it difficult to observe chiral logarithm  
behavior from data.

\section{Conclusions and future plans}

We have reported the results for the meson spectrum,
light quark masses and other physical quantities obtained from our
simulations in $N_f=2+1$ QCD with Wilson-clover action. 
We find that the meson spectrum is consistent with experiment, and
the quark masses are smaller compared to often quoted phenomenological 
values. 

We regard the present result as a first step toward a fully 
satisfactory $N_f=2+1$ simulation. 
A major point to improve is the control of systematic errors due to 
quark masses still large compared to those in Nature.  Larger volumes will 
also be needed for baryons, and even for mesons as quark masses are reduced. 
We hope to overcome these problems 
with the PACS-CS project \cite{Aoki:2005-PACSCS},
using the improved algorithm
provided by the domain decomposition idea~\cite{Luscher:2003} and 
the cluster computer PACS-CS developed at University of Tsukuba.

\vspace*{-1mm}

\begin{acknowledgments}
This work is supported by 
the Epoch Making Simulation Projects of Earth Simulator Center,
the Large Scale Simulation Program No.132 (FY2005) of 
High Energy Accelerator Research Organization (KEK),
the Large Scale Simulation Projects of 
Academic Computing and Communications Center, University of Tsukuba,
Inter University Services of Super Computers of 
Information Technology Center, University of Tokyo,
Super Sinet Projects of National Institute of Informatics,
and also by the Grant-in-Aid of the Ministry of Education
(Nos. 13135216, 13135204, 15540251,
      16540228, \ 16740147, \ 17340066, \ 17540259, \ 17740171, \
      18104005, \ 18540250, \ 18740130,
      18740167).
\end{acknowledgments}

\vspace*{-1mm}


\end{document}